\documentclass[a4paper]{jpconf}
\usepackage{graphicx}

\begin{document}

\title{TeV Particle Astrophysics II: Summary comments\footnote{Research
supported in part by the U.S. Department of Energy under DE-FG02 91ER 40626}
}
\author{Thomas K. Gaisser } 
\address{
Bartol Research Institute and Department of Physics and Astronomy\\ 
University of Delaware,
Newark, DE 19716 USA}

\ead{tgaisser@bartol.udel.edu}

\begin{abstract}
A unifying theme of this conference was the use of different
approaches to understand astrophysical sources of energetic particles in the TeV range and above.
In this summary I review how
gamma-ray astronomy, neutrino astronomy and (to some extent) gravitational wave astronomy
provide complementary avenues to understanding the origin and role of high-energy particles
in energetic astrophysical sources.
\end{abstract}

\vspace{-3pc}

\begin{figure}[h]
\vspace{0pc}
\hspace{.2pc}\begin{minipage}{19pc}
\begin{center}
\includegraphics[width=19pc,height=16pc]{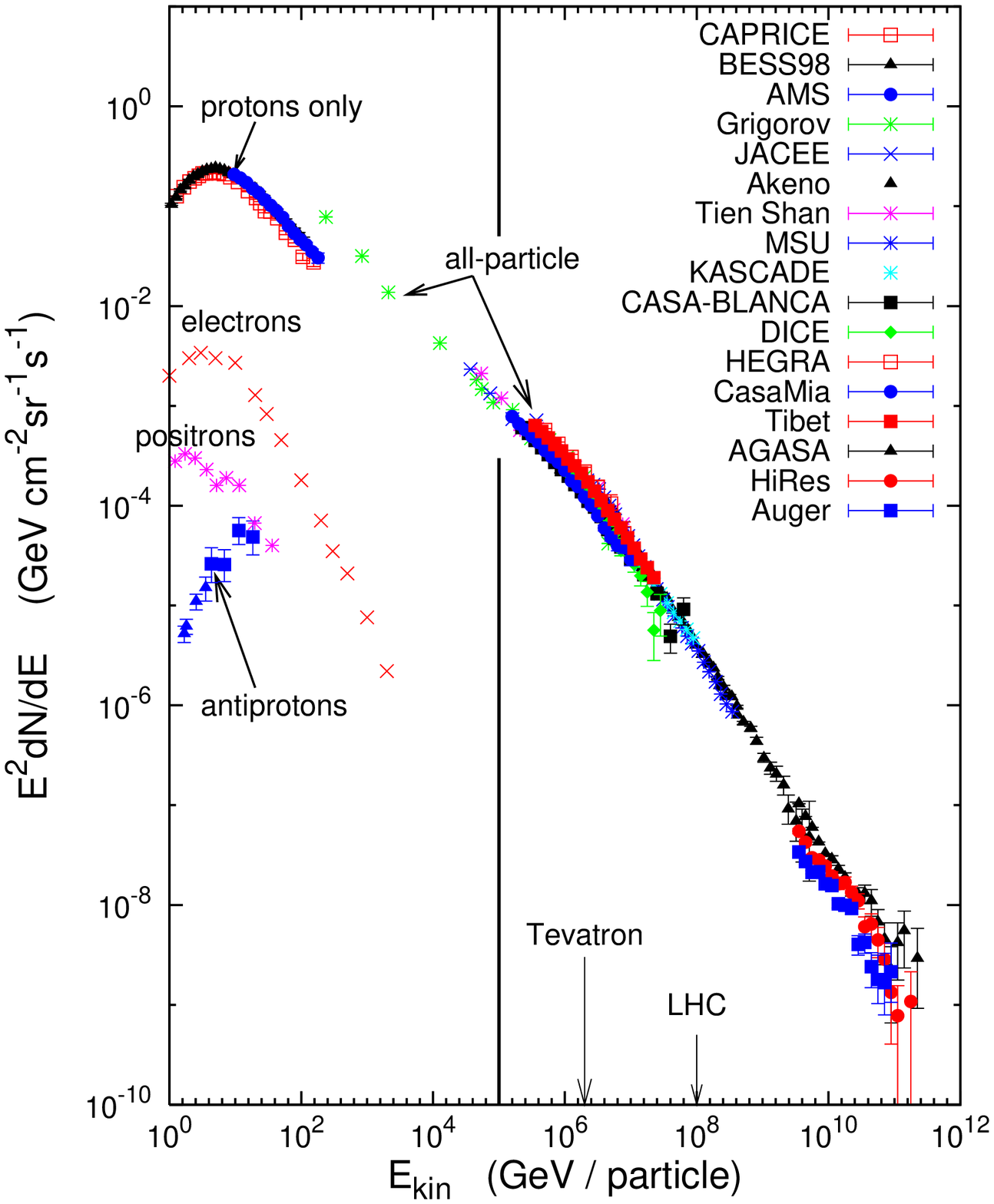}\vspace{-.1pc}
\caption{\label{fig1}The $\nu\,F(\nu)$ plot for cosmic rays.}
\end{center}
\end{minipage}\hspace{1pc}%
\vspace{-.1pc}\begin{minipage}{18pc}
\begin{center}
\includegraphics[width=18pc,height=12pc]{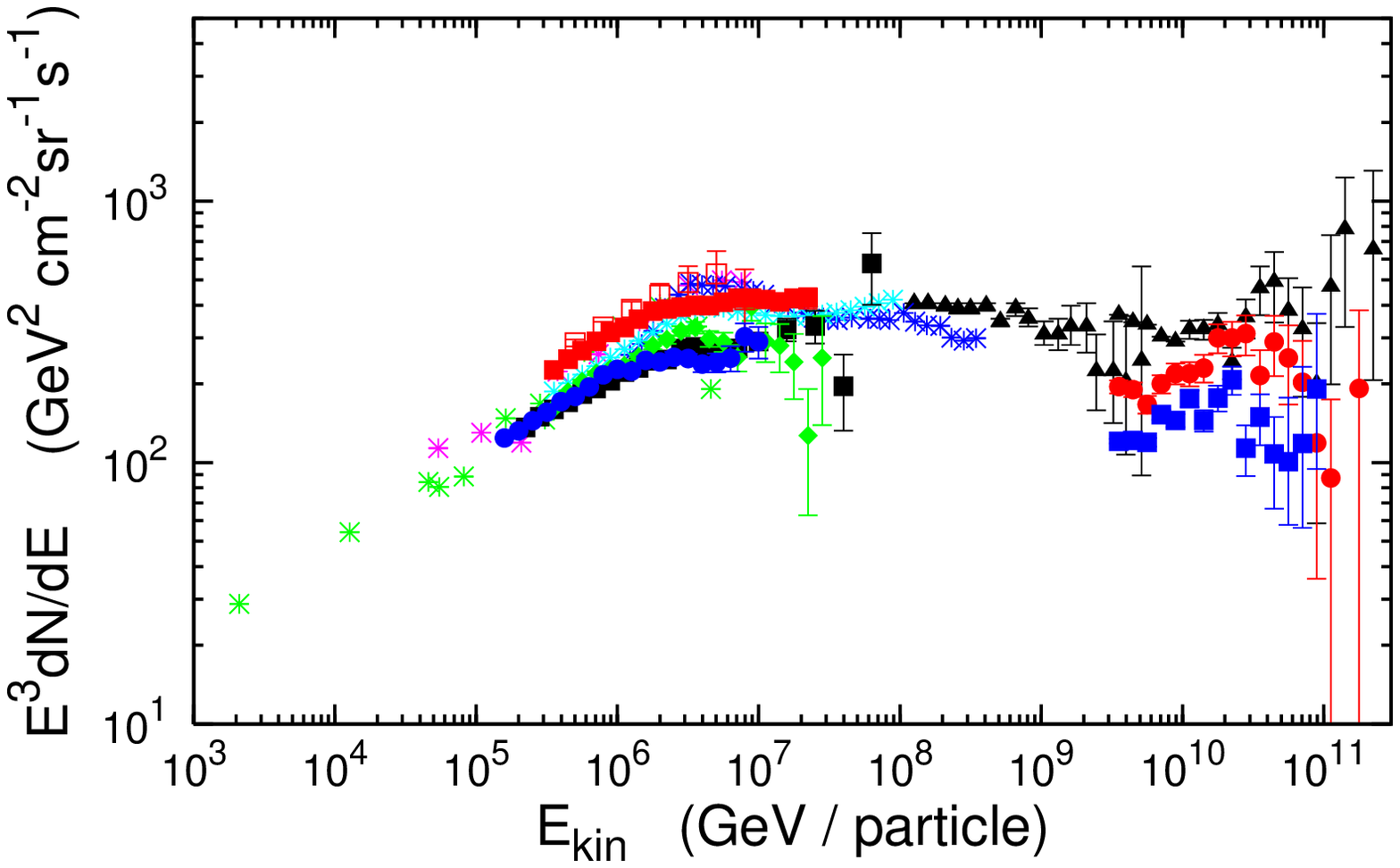}\vspace{.1pc}
\caption{\label{fig2a}The differential primary spectrum multiplied by $E^3$ (see text).}
\end{center}
\end{minipage} 
\end{figure}
\vspace{-1pc}

\section{Introduction: the cosmic-ray spectrum}

Figure~\ref{fig1} shows the energy spectrum of 
cosmic rays in the form of a "$\nu\,F(\nu)$" plot.  Multiplying
the differential energy spectrum by two powers of energy gives
a plot that shows the energy content per logarithmic interval of energy.  Most of the
energy is carried by particles with energies below 100 GeV.  In this energy
range, detailed measurements with magnetic spectrometers in balloons or
spacecraft above the atmosphere are possible.  Spectrometer measurements of protons, electrons, positrons and
antiprotons are shown on the plot.  Not shown are the measurements of 
individual nuclear species, such as lithium, beryllium, boron, carbon,
oxygen etc. from which we have the basic information about the lifetime
of cosmic-rays in our galaxy and the power needed to maintain them in equilibrium.
In his talk in parallel session IV, Swordy summarized the
current status of this field~\cite{Swordy}.

Above several hundred GeV, direct measurements are typically
made with calorimeters.  The charges of individual nuclei can still be determined,
but the systematics of the energy determination is not as good.  Above about $100$~TeV
(marked by the vertical line in Fig.~\ref{fig1}), the flux becomes so low that large air shower
detectors on the ground must be used.  From the ground only secondary cascades are measured,
so only indirect and very approximate measurements of the primary composition are 
possible.\footnote{Wakely's talk~\cite{Wakely} describes a method~\cite{direct-cherenkov}
that may be used to measure primary cosmic rays in the PeV range from the ground
by detecting direct Cherenkov light before the primary nuclei disintegrate.  The method works
best if Z is not too small.}  With an enclosed area of $3000$~km$^2$
the Auger Project~\cite{Auger} currently has the largest acceptance of any cosmic-ray detector.
First results with the partially completed Auger detector were presented at the 2005
Cosmic Ray Conference~\cite{Sommers} and are included in Fig.~\ref{fig1}.  
In his talk at this conference~\cite{Cronin},
Cronin pointed out that (except possibly at the very highest energies) the spectra measured
by AGASA~\cite{AGASA}, Hi-Res~\cite{HiRes} and Auger~\cite{Sommers} can be brought into
close agreement by a systematic shift in energy of the order of 25\%.

Figure~\ref{fig2a} shows the same data as in Fig.~\ref{fig1} in the form 
conventionally used for discussion of the spectrum of ultra-high energy cosmic rays.
Such a plot was shown and discussed by Blasi~\cite{Blasi}
in his talk at this conference.  He noted several features in the spectrum:
\begin{itemize}
\item The knee around $3\times 10^{15}$~eV,
\item A second knee above $10^{17}$~eV,
\item A dip just below $10^{19}$~eV and
\item A GZK feature in the decade above $10^{19}$~eV.
\end{itemize}
Except for the ``second knee", these features are clearly visible in
the data collected in Fig.~\ref{fig2a}.  The first ``knee" is a
steepening of the spectrum from a differential index of $\approx -2.7$
below $10^{15}$~eV to $\approx -3$ for $E>10^{16}$~eV.

Peters showed long ago~\cite{Peters} that if the knee in the spectrum represents the
transition from a low energy population of cosmic rays to a higher energy population
with a harder spectrum, then one should see a characteristic sequence of changes
in the elemental composition.  
The underlying assumption of his argument is that the spectral change
is the result of a physical effect that depends on magnetic rigidity of the 
cosmic rays, which is the case for many models of propagation and acceleration.
For example, for  shock acceleration by supernova remnants, when the gyroradius
of a particle exceeds the size of the remnant, it will escape from the 
acceleration region.  Thus the criterion that determines the maximum energy
is that the gyroradius of a particle should not exceed the radius of the
remnant, $R_{\rm SN}$~\cite{Bell}.  This leads to   
\begin{equation}
E_{\rm max}\;=\;\xi\beta\times Z\,e\,B\,R_{\rm SN},
\end{equation}
where $Z\,e$ is the charge of a nucleus of total energy $E$, $B$ the
magnetic field strength in the acceleration region and
$\xi\beta$ the supernova shock velocity reduced by a numerical factor $\xi$.
In this sequence, first protons would decrease in abundance, then
helium ($Z=2$) then the CNO group, etc.  

Several air-shower experiments
(e.g. Refs.~\cite{EASTOP,SPAM} and references therein) show
a general trend of increasing mass through the knee region,
but the KASCADE~\cite{KASCADE,Haungs} experiment is the first to see evidence of
the sequence of steepening of individual spectra predicted by Peters.
The KASCADE result, in common with all indirect determinations of
the composition, is limited by uncertainties in the models of hadronic
interactions used to interpret the data.  Thus the fractions
of the different groups of nuclei are less well established than
the energy-dependence of the average mass, which becomes heavier
through the knee region.  Current experiments become statistically
limited below $10^{17}~eV$.  This is just the energy range one would
like to explore in more detail to look for signatures in composition of
a transition to a new (presumably extra-galactic) population of
particles at higher energy.

The location in energy of a transition from Galactic to extra-Galactic
cosmic rays is a question of long standing.  In the past, the
dip was sometimes interpreted as the signature of the transition
to an extra-galactic population~\cite{FlyEye}.  More recent data from the HiRes
experiment~\cite{Sokolsky} seem to suggest a transition from mostly heavy
nuclei (interpreted as the end of galactic cosmic rays) to mostly light nuclei
(possibly extra-galactic) an order of magnitude or more lower in energy.
This result would be consistent with the interpretation by Berezinsky et al.~\cite{Berezinsky}
of the dip as a consequence of energy losses by protons producing electron-positron 
pairs in interaction with the cosmic background radiation.

If the extra-galactic component includes too large a fraction of heavy nuclei,
however, the dip due to pair production is wiped out.  A recent paper,~\cite{Allard}
reported on at this conference by Busca~\cite{Busca}, contrasts two models of
extra-galactic cosmic rays.  One is mostly protons, in which case the
dip is explained by pair production and the transition to extra-galactic
cosmic rays occurs at relatively low energy ($<10^{18}$~eV).  The other
postulates a large fraction of heavy nuclei and interprets the dip
as the signature of the onset of the extra-galactic component.
In the latter case, one probably needs a special, high-energy galactic
component (called ``component B" by Hillas~\cite{hillas06}) to fill
in between conventional supernova origin and the extra-galactic
component at ultra-high energy.

Composition in the intermediate region between the knee and the dip is
an important unanswered question at present.  First spectrum measurements
from KASCADE-Grande extending to $10^{18}$~eV
were shown at this conference~\cite{Haungs}, but not
yet interpreted in terms of composition.  The IceCube experiment~\cite{performance05}, currently
under construction at the South Pole, with its surface component IceTop
will also be sensitive to primary composition up to $10^{18}$~eV.  A low-energy
extension of the telescope array is contemplated~\cite{TALE} to explore this
energy region, and there are other possibilities~\cite{Tunka}, including
use of the radio technique to detect showers by their geo-synchrotron
radiation~\cite{Falcke,LOPES}.

\section{Multi-messenger astronomy}
The term ``multi-messenger astronomy" was first used, as far as I can ascertain,
by Steve Barwick during one of the committee meetings of the panel on
Particle, Nuclear and Gravitational-Wave Astrophysics in the last decadal
survey~\cite{decadal}.  What was then to some extent a rhetorical
device in our report is now coming closer to reality as neutrino telescopes
approach the sensitivity needed to see neutrinos from established and likely
TeV gamma-ray sources.  Ideas for coordinating searches between gamma-ray
and neutrino telescopes are now being proposed~\cite{Bernardini,Resconi}.
Coincident observation of a gamma-ray burst by LIGO is also a possibility
in the near future~\cite{Cornish}.

In his review of ground-based gamma-ray astronomy, Eckart Lorenz~\cite{Lorenz} showed
a sky plot with well over 30 $\ge$~TeV~$\gamma$-ray sources, including 
both galactic and extra-galactic objects.  His map showed only a fraction of the
sources newly discovered by H.E.S.S.~\cite{HESS} in the inner galaxy~\cite{Horns}.
Several of these are supernova remnants, some of which are clearly
better fit by hadronic models than by synchrotron-self-Compton models.  An example
is RXJ1713.7-3946, first reported in TeV by CANGAROO~\cite{CANGAROO}.  The results
from H.E.S.S. observing the southern sky are making a breakthrough in VHE gamma-ray
astronomy.  VERITAS is now coming on line with
four telescopes~\cite{Krennrich} and will have similar sensitivity for the northern sky.
The recent observation by MILAGRO of an excess from the Cygnus region~\cite{Goodman} of the
galactic plane will be particularly interesting to investigate with VERITAS.

\begin{figure}[h]
\vspace{-.5pc}
\hspace{.2pc}\begin{minipage}{18pc}
\begin{center}
\includegraphics[width=18pc,height=13pc]{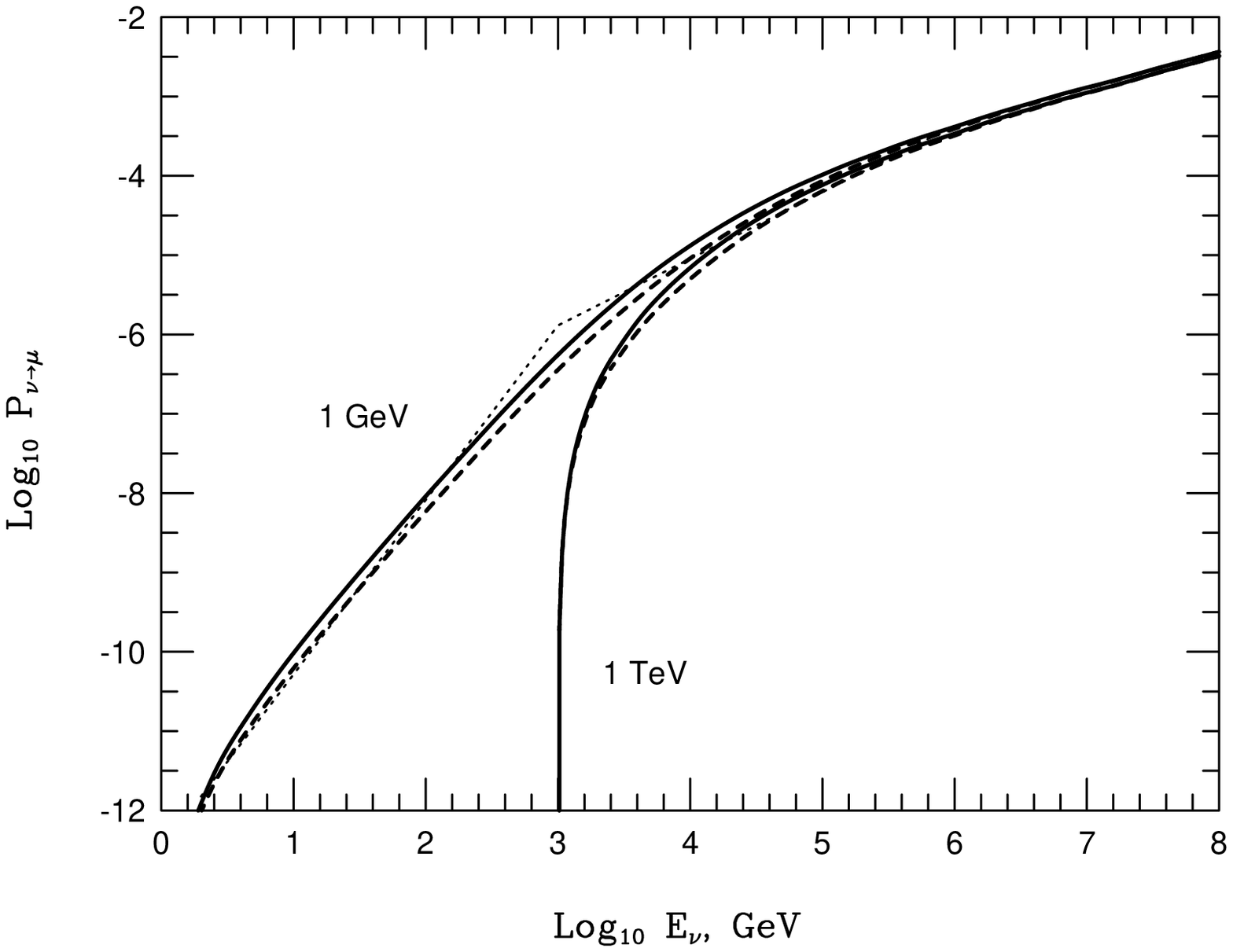}\vspace{-.5pc}
\caption{\label{fig3}The probability $P_\nu(E_\nu)$ for detecting a muon neutrino (solid)
or anti-neutrino (dashed) by the charged current interaction channel~\cite{GHS}.
(See text.)}
\end{center}
\end{minipage}\hspace{2pc}%
\vspace{-1.5pc}\begin{minipage}{18pc}
\begin{center}
\includegraphics[width=18pc,height=12pc]{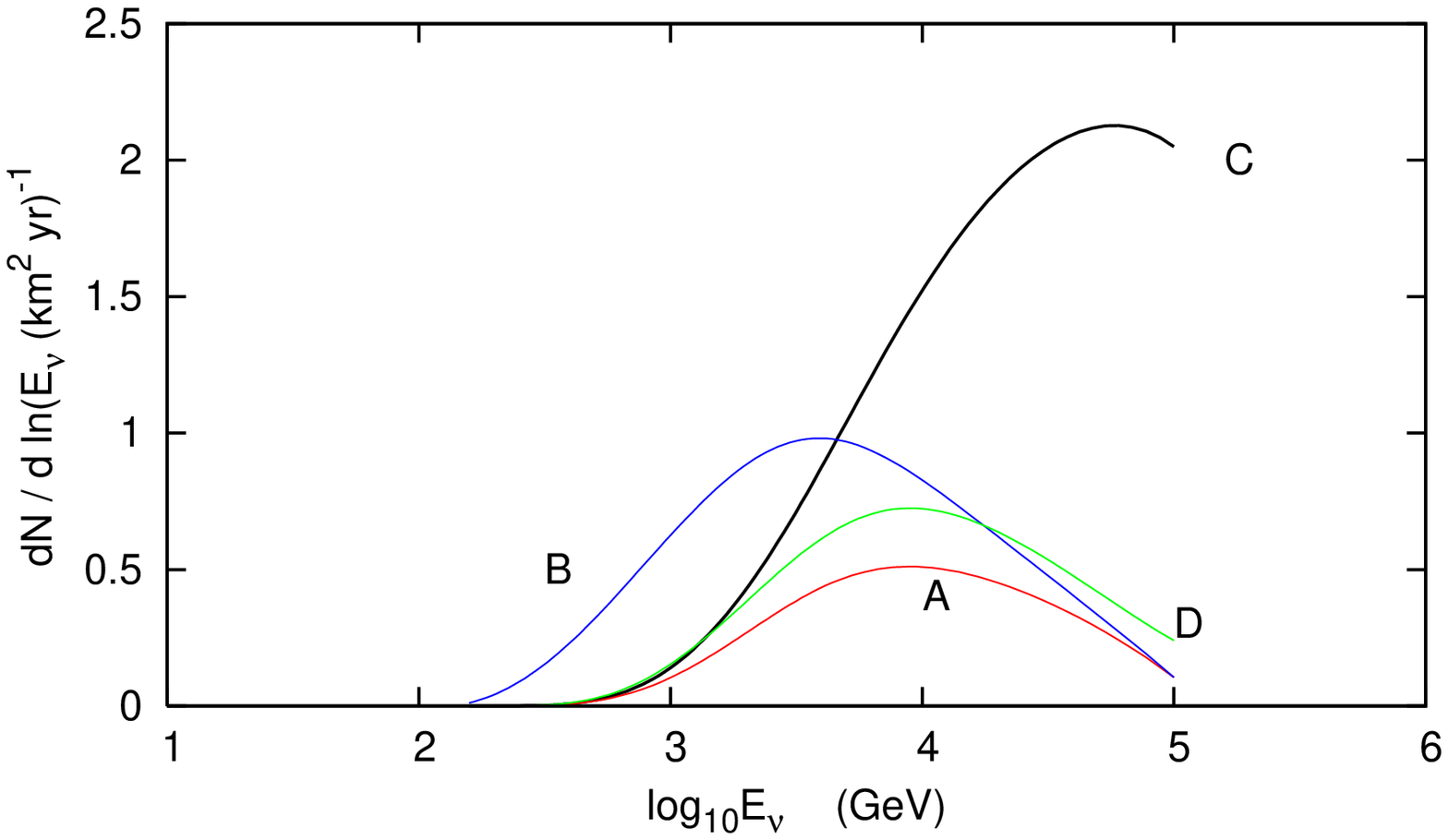}\vspace{-.1pc}
\caption{\label{fig4}The differential response function for neutrino signals.
Curves A, B, C and D correspond to different sources and assumptions as described
in the text.}
\end{center}
\end{minipage} 
\end{figure}
\vspace{1pc}

In his survey of astrophysical sources of neutrinos and expected rates~\cite{Dermer}, Dermer
stated a rule of thumb for detectability of neutrino sources in kilometer scale neutrino
telescopes.  An energy fluence of $10^{-4}$~ergs/cm$^2$ is the criterion for detectability.  
The argument, which assumes that muon neutrinos are the favored channel because of the large muon
range, is illustrated in Figs.~\ref{fig3},\ref{fig4}.  The signal is the convolution
of the neutrino flux, $\phi_\nu(E_\nu)$, with the probability $P_\nu(E_\nu)$
that a neutrino on a trajectory toward the detector
will interact to produce a muon that reaches the detector with enough energy to be detected.
In general, the detection efficiency will be an increasing function of energy.  Figure~\ref{fig3}
from Ref.~\cite{GHS} shows two extreme examples, step function thresholds at 1 GeV and 1 TeV.

Figure~\ref{fig4} shows the response function, $E_\nu\,\phi_\nu(E_\nu)\times P_\nu(E_\nu)$,
for several examples.  The response function is essentially dN$_{\rm signal}$/d{\cal l}n($E_\nu$),
so that equal areas on the plot correspond to equal contributions to the signal.  Curves A and B
refer to the spectral shape of RXJ1713.7-3946 as measured by H.E.S.S.~\cite{HESSRXJ1713},
which shows evidence of steepening, consistent with an exponential cutoff above 10 TeV.
The neutrino flux is normalized assuming all observed gamma-rays are from $\pi^0$-decay
without absorption in the source.  The flux at Earth is reduced by a kinematic factor
($\approx 0.85$) that relates $\pi^0$ photons to $\nu_\mu + \bar{\nu}_\mu$ from the charged 
$\pi\rightarrow\mu\rightarrow e$ decay chain and by a factor of two to account for 
oscillations.  Absorption of neutrinos in the Earth is unimportant below 100 TeV. 
The integrated energy fluence of $\nu_\mu + \bar{\nu}_\mu$ at Earth with these
assumptions is $\approx 2\times 10^{-4}$~erg/cm$^2$.  Curve
A corresponds to a requirement $E_\mu>1$~TeV at the detector, while the threshold for
curve B is taken as $100$~GeV.  These threshold are meant to bracket
the true threshold for a kilometer-scale neutrino detector.  
The corresponding range of signals is $\sim 2$ to $4$ events per 
square kilometer per year.  These numbers are comparable to the atmospheric
background within a cone of radius $1^o$.

Curve C is the response function for a
neutrino spectrum with the same normalization as A at $1$~TeV but with a
hard differential spectral index of $-2$ and cutoff above 1 PeV.  
Such a situation is conceivable
(though unlikely) if the gamma-ray cutoff is due to absorption at the source
in a way that does not affect the neutrinos.  Curve C also illustrates how
detectors respond to different
neutrino spectra.  For example, for
power-law spectra with differential indices of -3, -2 and -1, Becker~\cite{Becker}
showed that the response for AMANDA peaks respectively at $0.5$, $50$ and $10^5$~TeV.

Curve D in Fig.~\ref{fig4} shows the integrated diffuse flux corresponding to 
the MILAGRO gamma-ray spectrum for the Cygnus region.  Averaged over a region
of $\pm 3^o$ in galactic latitude and 20 degrees in galactic longitude,
the integral gamma-ray intensity is $4\times 10^{-13}$~(cm$^2$s)$^{-1}$
for $E_\gamma>12.5$~TeV with an assumed differential spectral index of $-2.6$.
The corresponding upward muon signal would be $\sim 3$ per km$^2$yr, under
the same assumptions described above for relating neutrinos at Earth
to the observed spectrum of photons.  Because of the diffuse nature of the source,
however, such a low signal would be hidden in the atmospheric background.

\section{Concluding comments}
In his summary~\cite{Dermer}, Dermer listed blazar active galaxies
(``flat spectrum radio quasars, not BL Lacs") and gamma-ray bursts as the most promising
sources for kilometer-scale neutrino detectors.  Finding the first signals from such sources,
as well as from the kinds of Galactic sources mentioned above,
by detecting upward neutrinos in the TeV-PeV range is the primary goal of
IceCube, now under construction in Antarctica~\cite{performance,Kael}, and of planned 
kilometer scale detectors in the Mediterranean~\cite{Coyle}.
The ability to distinguish between single muons
and muon bundles with Antares~\cite{Bouwhuis} illustrates the merits of water as a detection
medium.  Good progress with construction of IceCube~\cite{Kael} and the successful
performance of the first strings~\cite{performance}
show that the Antarctic ice sheet is also a suitable detection medium,
The solid surface is convenient for deployment and readily accommodates
the IceTop air shower array as an integral component of IceCube.
The two large neutrino telescope projects
are complementary in technique as well as sky coverage.  

In view of the
low level of expected signals, it is important to develop techniques for
multi-messenger astronomy to help discriminate signals from background.
In this respect, searching for neutrinos associated with identified gamma-ray
burst events is attractive.  More generally, finding a statistically proper
way to define and search for coincidences between gamma-ray activity and
neutrino signals should be helpful~\cite{Bernardini,Resconi}.  Triple
coincidences with gravitational wave detectors as well would be even better.

Yoshida~\cite{Yoshida} discussed
GZK neutrinos as a probe of models of cosmic evolution and sources of ultra-high energy
cosmic rays.  Weiler~\cite{Weiler} discussed the possibility of using precise
measurements of neutrino flavor ratios to probe conditions in the acceleration region
as well as looking for novel intrinsic neutrino properties.  He also discussed
using horizontal events to probe the neutrino cross section at ultra-high energy.
To fully exploit the potential of UHE neutrino astronomy appears to require
new techniques capable of seeing larger target volumes with more sparsely
instrumented arrays.  Ideas for radio and acoustic detection of neutrinos 
were reviewed by Varner~\cite{Varner} and Thompson~\cite{Thompson} respectively.
These techniques have higher energy thresholds (particularly the acoustic), so
finding a way to cross-calibrate with other techniques is important and is
a recognized goal.

A better understanding of hadronic interactions is important for
interpreting the results of air shower experiments as well as for obtaining better estimates
of the atmospheric muon and neutrino background in neutrino telescopes.
Catanesi~\cite{Catanesi} described new measurements with fixed target experiments
at CERN which should provide the basis for improving treatment of the low-energy
interactions in calculations of air showers as well as the neutrino flux calculations.
Processes of interest for particle physics in the LHC era have cross sections
in the microbarn range and below~\cite{Han}.  As a consequence, despite their somewhat higher reach
in energy, air shower experiments are unlikely to make important new discoveries
in particle physics.  On the other hand, any information about the fragmentation
region of hadronic interactions forthcoming from forward detectors at the LHC
will be of great value for interpreting ultra-high energy cosmic-ray cascades 
to determine primary composition.

The most profound problems addressed at the conference concern dark matter and the 
early universe.  The WMAP data are described well
by the Lambda Cold Dark Matter model~\cite{Komatsu}, which requires both dark energy
and cold dark matter.  Cabrera~\cite{Cabrera} and Aprile~\cite{Aprile} 
reviewed direct searches for dark matter.  Ullio~\cite{Ullio} reviewed the many
ways hints of dark matter might show up indirectly in spectra of particles and
radiation, the most striking of which would be a diffuse gamma-ray line, which would
give a direct measure of the WIMP mass.  Among the viable possibilities is
a neutrino signal from WIMP capture and annihilation in the Sun large enough to be
detected in AMANDA/IceCube.



\end{document}